\documentclass{aastex}
\usepackage{spr-astr-addons}
\usepackage{url}

\RequirePackage{color}
\begin{document}
%
\title{Vertically Self-Gravitating ADAFs in the Presence of Toroidal Magnetic Field}


\author{A. Mosallanezhad \altaffilmark{1}}
\and
\author{S. Abbassi\altaffilmark{1,2}}
\affil{amin.mosallanezhad@gmail.com,\ abbassi@ipm.ir}
\and
\author{M. Shadmehri \altaffilmark{3}}
\affil{m.shadmehri@gu.ac.ir}
\and
\author{ J. Ghanbari\altaffilmark{4,5}}
\affil{ghanbari@ferdowsi.um.ac.ir}
\altaffiltext{1}{School of Physics, Damghan University of Basic Sciences, Damghan, 36715-364, Iran}
\altaffiltext{2}{School of Astronomy, Institute for Research in Fundamental Sciences (IPM), Tehran, 19395-5531, Iran}
\altaffiltext{3}{Department of Physics, Faculty of Science, Golestan University, Basij Square, Gorgan, Iran}
\altaffiltext{4}{Department of Physics, School of Sciences, Ferdowsi University of Mashhad, P.O.Box 91775-1436, Mashhad, Iran}
\altaffiltext{5}{Khayam Institute of Higher Education, P.O.Box 918974-7178, Mashhad, Iran}

\begin{abstract}
Force due to the  self-gravity of the disc in the vertical direction is considered to study its possible effects on the structure of a magnetized advection-dominated accretion disc. We  present steady-sate self similar solutions for the dynamical structure of such a type of the accretion flows. Our solutions imply reduced thickness of the disc because of the self-gravity. It also imply that the thickness of the disc will increase by adding the magnetic field strength.
\end{abstract}


\section{INTRODUCTION}
There has been rapidly progress over the past three decades towards  a better  understanding of the accretion processes in astrophysics, in particular
accretion discs around compact objects or even black holes (see reviews by Narayan, Mahadevan \& Quataert 1998; Kato, Fukue \& Mineshige 2008). Black hole accretion is one of the most
important ingredient when considering astrophysical scenarios of galaxy/quasar formation.  Current belief is that there
are two sites where an accretion disc around a black hole can be found: In close binary systems called X-ray binaries (XBs), and at the center of the galaxies.
The birth of modern accretion disc theory is traditionally attributed to the original model presented by Shakura \& Sanyev (1973).
This standard geometrically thin, optically thick accretion disc model (SSD) can successfully explain most of observational features in active galactic nuclei (AGNs)
and X-ray binaries. In the standard thin disc model, the motion of matter in the accretion disc is nearly Keplerian, and the viscous heat in the disc is radiated away locally.
An alternative accretion disc model, namely, the advection-dominated accretion flows (ADAFs),
was suggested for the black holes accreting at very low rates (Ichimaru 1977, Narayan \& Yi 1994).

In some of the scenarios of structure formation in astrophysics, in particular those related to the formation of the stars
or galaxies, self-gravity of the system may play a vital role. Formation of an accretion disc, as one stage of the structure formation,
is an important part of any theory in this field. However,  in the standard accretion disc model, the effect of self-gravity in the vertical
or radial direction of the disc is neglected for simplicity and the disc is supported in the vertical direction only by the thermal pressure.
Although in some of the accreting systems it is a reasonable assumption, there are situations, in which one can hardly neglect self-gravity of the disc itself.
 According to the current theories, self-gravity of the disc not only can contribute in generating turbulence inside the disc but its force in the radial and the vertical
directions would modify the angular velocity and and the vertical
scale-hight, respectively Duschl et al. 2000. Therefore, it is not
an easy task to include self-gravity of  a disc in a self-consistent
approach. In fact, the theory of self-gravitating accretion discs is
less developed. Early numerical work of self-gravitating accretion
discs began with N-body
    modeling (Cassen \& Moosman 1981; Toomley; Cassen \& Stein-Cameron 1991). Although great progress has been made in recent
years in increasingly sophisticated numerical accretion disc
simulations, simple analytic disc models still are the only
accessible way of making direct link between the theory and
observations. The theoretical treatment can estimate the spectra and
other observational features of the accretion powered objects. To
solve the nonlinear equations of the self-gravitating accretion
discs, sometimes the technique of self-similarity is useful. Some of
the astrophysical systems often attain self-similar limits for a
wide range of the initial conditions. Moreover, the self-similar
properties  allows us to investigate properties of the solutions in
arbitrary details, without any of the associated difficulties of
numerical magneto-hydrodynamics. Several classes of the self-similar
solutions for the structure of self-gravitating accretion discs have
been studied by now (e.g., Paczynski 1978; Mineshige \& Umemura
1996;  Shadmehri 2004). Some of them are presented for the standard
accretion discs (e.g., Paczynski 1978), while the others are
describing ADAFs including self-gravity (e.g., Mineshige \& Umemura
1996; Shadmehri 2004). Abramowicz et al. 1984 examined a
number of issues related disc's self-gravity in a steady state
accretion disc, especially in regard to the effect of disc
self-gravity on the disc topology and disc dimensions. Bu, Yuan \&
Xie (2009) have proposed a self-similar solution of a magnetized
ADAFs without self-gravity.  Mineshige \& Umemura (1996) presented
a set of self-similar solutions for steady-state structure of ADAFs
considering gravitational force of the disc in the radial direction.
However, such solutions are hardly applicable because the
gravitational force of the disc due to its self-gravity in ADAfs is
generally smaller than the radial force of the central object.
However, the gravitational force of the disc in the vertical
direction can be a fraction of the vertical component of
gravitational force of the central object, at least at the outer
part of ADAFs. Then, as we will show, the structure of the disc
would modify because of the self-gravity of the disc itself. We
think, our solutions are applicable to the outer parts of ADAFs in
X-ray binaries.

Although we did not apply our solution to a specific
astronomical system, we believe the solutions are applicable to the
discs with an inner hot part surrounded by an outer cold part. In
the most cases, one can hardly recognize the transition region. For
example, the outer regions of the X-ray binaries are SSD, though
their inner parts resemble to the hot accretion flows (e.g., Esin et
al. 1997). Our solutions are suitable for a region where the disc is
self-gravitating but the advection of the dissipated energy is not
negligible. Also, the discs of the low-luminosity AGNs (e.g., Ho
2008) may have a similar situation and they do not show clear
transition from SSD to ADAF. The solutions are also applicable to
Slim discs, and thus have its potential applications in ULXs and
NLS1s (Mineshige et al. 2000; Watarai et al. 2000).  The next step
for future studies is to relax our simplifying assumptions, in
particular similarity method, and solve the relevant equations for
one of the mentioned systems and in doing so, our similarity
solutions will guide us.

\section{The Basic Equations}

We are interested in analyzing the dynamical behavior of a magnetized ADAF when the force due to the self-gravity of the disc
in the vertical direction is considered. But self-gravitational force in the radial direction is neglected because ADAFs do not
extend radially so much. We  consider only toroidal component of the magnetic field. We suppose that the gaseous disc is rotating
around a compact object of mass $ M_* $. Thus, for  a steady axisymmetric accretion flow, i.e. $\partial/ \partial t=\partial / \partial \phi=0$,
we can write the standard equations in the cylindrical coordinates  $ (r,\phi,z) $. We vertically integrate the flow equations and so, all the
physical variables become only functions of the radial distance $r$. We also neglect  the relativistic effects and  Newtonian gravity in radial
direction is considered. The disc is supposed to be turbulent and possesses an effective turbulent viscosity $\bf \nu$.  As for the energy
conservation, we assume   the generated energy due to viscosity dissipation is balanced by the radiation and advection cooling (e.g., Narayan \& Yi 1994).

The equation of continuity gives,
\begin{equation}
\frac{1}{r}\frac{d}{d r}(r\Sigma
v_{r})=2 \dot{\rho}H
\end{equation}
where $v_{r}$ is the accretion velocity (i.e., $v_{r}<0$) and $ \dot{\rho} $ denotes the mass-lose rate  per unit volume due to the wind or outflow, $ H $ is the disc half-thickness,
and $\Sigma$ is the surface density. We can also write $ \Sigma=2 \rho H $.

The equation of motion in the radial direction is
\begin{equation}
v_r\frac{d v_r}{d r}=\frac{v_\varphi^{2}}{r}-\frac{G
M_{\ast}}{r^{2}}-\frac{1}{\Sigma}\frac{d}{dr}(\Sigma
c_s^{2})-\frac{c_A^{2}}{r}-\frac{1}{2\Sigma}\frac{d}{dr}(\Sigma
c_A^{2}),
\end{equation}
where $v_{\varphi}, c_s$ and $c_A$ are the rotational velocity of the disc, sound and Alfven velocities of the gas, respectively.
Sound speed and the Alfven velocity are defined as $c_{\rm s}^2=p_{\rm gas}/ \rho$ and  $c_{\rm A}^2=B_{\varphi}^2 / 4\pi\rho=2p_{\rm mag}/ \rho$ where $p_{gas}$
and $p_{mag}$ are the gas and the magnetic pressures, respectively.

The vertically integrated angular momentum equation becomes
\begin{equation}
r\Sigma v_r
\frac{d}{dr}(rv_\varphi)=\frac{d}{dr}(r^{3}\nu\Sigma\frac{d\Omega}{dr})
\end{equation}
where $ \Omega = v_\phi / r $ is the angular velocity.
Also $\nu$ represents the kinematic viscosity coefficient and we assume (Shakura \& Sunyaev 1973),
\begin{equation}
\nu=\alpha c_s H
\end{equation}
where $\alpha$ is a  nondimensional parameter less than unity.\\

The goal of this {study} is to investigate the effect of
self-gravity of the disc on  its structure. One can estimate the
importance of self-gravity of the disc by comparing the
contributions to the local gravitational acceleration in the
vertical direction by both the central object and the disc itself.
The vertical acceleration at the disc surface due to its
self-gravity is $2\pi G\Sigma$ and by the central object is
$GM_{*}h/r^3$. Thus,  force due to the self-gravity of the disc in
vertical direction is  dominated if:
\begin{equation}
\frac{M_d}{M_*}\sim\frac{\pi r^2 \Sigma}{M_*}>\frac{1}{2}\frac{h}{r}
\end{equation}
where $M_d$ is the mass enclose the disc within a radius $r$. For ADAFs the typical values of $h/r$ is around 1. Since the enclosed mass $M_d$ is an increasing function of the radial distance
$r$, the effect of self-gravity becomes significant when the disc is thick. Here we consider keplerian selfgravitating (KSG) disks in which selfgravity
is significant only in the vertical direction and which satisfy the constraint $ (1/2)(h/r)M_*\leq M_d(r)\leq M_* $. Compared to the standard ADAF solutions (Narayan \& Yi 1994; Akizuki \& Fukue 2006),
the KSG disks require modification of the equation of hydrostatic support in the direction perpendicular to the disk. Thus, while in the standard ADAF solutions the local vertical pressure gradient is balanced by the
z component of the gravitational force due to the central object, in the SG case we have balance between two local forces, namely the pressure force and the gravitational force due to the disk’s local mass.
Also in the KSG case, in the radial direction centrifugal forces are still balanced by gravity from a central mass (Keplerian approximation) .

For a self-gravitating disc, the hydrostatic equilibrium in the vertical direction yields (e.g., Paczyncki 1978, Duschl et al. 2000)
\begin{equation}\label{eqn:P_c}
P_c=\pi G \Sigma^2
\end{equation}
 where $ P_c $ is the pressure in the central plane $ (z=0) $ and $ G $ is the gravitational constant. Since details of the thermodynamics in the vertical direction
 are not considered in our model, we shall assume the disc to be isothermal in the vertical direction.

For a magnetized Keplerian self-gravitating disc, we have
\begin{equation}\label{eqn:C_s}
\frac{P_c}{\bar{\rho}}=c_s^{2}[1+\frac{1}{2}(\frac{c_A^{2}}{c_s^{2}})^2]=c_s^{2}(1+\beta)
\end{equation}
where $ \bar{\rho}=\frac{\Sigma}{2H} $ is a vertically averaged mass
density and
$\beta=\frac{P_{mag}}{P_{gas}}=\frac{1}{2}(\frac{c_A}{c_s})^{2}$,
where this parameter shows the important of magnetic field pressure
in comparison to the gas pressure.  We will take $\beta$ as an input
parameter of our model and its effect on the structure of the disc
is studied for different values of $\beta$. Having Equations
\ref{eqn:P_c} and \ref{eqn:C_s}, we can write
\begin{equation}
H=\frac{(1+\beta)c_s^{2}}{2 \pi G \Sigma}.
\end{equation}

Also, the energy equation becomes
\begin{equation}
\frac{\Sigma
v_r}{\gamma-1}\frac{dc_s^{2}}{dr}+\frac{c_s^{2}}{r}\frac{d}{dr}(rv_r)=f
\nu r^2(\frac{d\Omega}{dr})^{2}
\end{equation}
where $ \gamma $ is the ratio of specific heats. The advection parameter $ f $ measures the degree to which the flow is advected-dominated (Narayan \& Yi 1994), and it is supposed to be constant.

Induction equation with the field scape can be written as (Akizuki \& Fukue 2006)
\begin{equation}
\frac{d}{dr}(V_r B_\varphi)=\dot{B_\varphi},
\end{equation}
where $\dot{B_\varphi}$ is the field scaping/creating rate due to
the magnetic instability or the dynamo effect. We can rewrite this equation as
\begin{equation}
v_r
\frac{dc_A^{2}}{dr}+c_A^{2}\frac{dv_r}{dr}-\frac{c_A^{2}v_r}{r}=2c_A^{2}\frac{\dot{B_{\phi}}}{B_{\phi}}-c_A^{2}\frac{2
\dot{\rho}H}{\Sigma}.
\end{equation}

Now, we have a set of ordinary differential equations that describe the dynamical behavior of a magnetized ADAF flow including self-gravity of the disc.
Solutions of these equations give us the dynamical behavior of the disc, which depends on the viscosity, magnetic field strength, self-gravity and advection rate of energy transport.

\begin{figure}[t]
\input{epsf}
\epsfxsize=3.8in \epsfysize=3.2in
\centerline{\epsffile{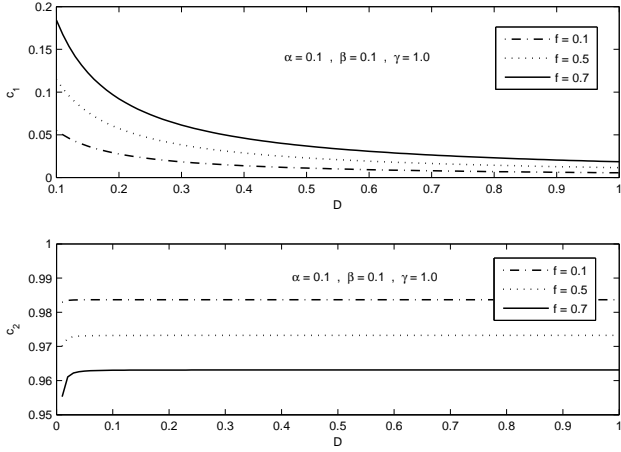}}
 \caption{These plots show behavior of the  coefficients $ C_1 (-\frac{v_r}{\alpha v_k}) $ and $C_2 (\frac{v_{\varphi}}{v_k})$ versus  self-gravitating parameter $D$ for different  values of the advection parameter $f$. Here, the input parameters are  $ \beta=0.1 $, $ \alpha=0.1 $  and $ \gamma=1 $. }\label{c1-f_alpha}
\end{figure}

\begin{figure}[t]
\input{epsf}
\epsfxsize=3.8in \epsfysize=3.2in
\centerline{\epsffile{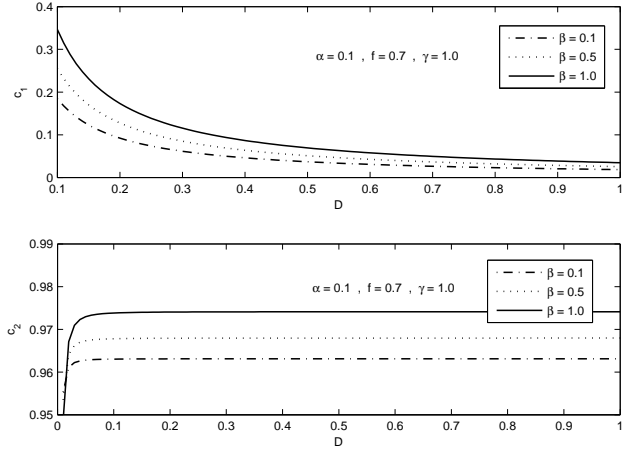}}
 \caption{These plots show behavior of the  coefficients $ C_1 (-\frac{v_r}{\alpha v_k}) $ and $C_2 (\frac{v_{\varphi}}{v_k})$ versus  self-gravitating parameter $D$ for different values of the magnetic field parameter $\beta$. Here, the input parameters are  $ f=0.7 $, $ \alpha=0.1 $  and $ \gamma=1 $. }\label{c1-f_beta}
\end{figure}

\section{Self-Similar Solutions}

Global behavior of the disc can not be described using similarity solutions,
because boundary conditions are not considered in this method.
But, as long as we are not interested in the solutions near the boundaries, similarity solutions describe
correctly, true and useful asymptotically  behavior of the flow in the intermediate regions.

We assume that the physical quantities can be expressed as a power law of the radial distance,i.e. $ r^{\nu} $, where $ \nu $ is determined by
substituting the similarity solutions into the main equations and solving the resulting algebraic equations. Therefore, we can write similarity solutions as
\begin{equation}
v_r(r)=-C_1 \alpha v_k(r),
\end{equation}
\begin{equation}
v_\varphi(r)=C_2 v_k(r),
\end{equation}
\begin{equation}
c_s^{2}(r)=C_3 v_k^{2}(r),
\end{equation}
\begin{equation}
c_A^{2}(r)=\frac{B_\varphi^{2}(r)}{4\pi\rho(r)}=2\beta
C_3v_k^{2}(r),
\end{equation}
where
\begin{equation}
v_k(r)=\sqrt{\frac{GM}{r}},
\end{equation}
and the parameters $C_1$,$C_2$ and $C_3$ will be determined from the main equations.

In addition, the surface density $ \Sigma $ is assumed to be a form of
\begin{equation}
\Sigma=\Sigma_0 r^s,
\end{equation}
where $ \Sigma_0 $ and $ s $ are constants. Then, in order for the self similar treatment to be valid, the mass-loss rate per unit volume and the escaping rate must have the following form,
\begin{equation}
\dot{\rho}=\dot{\rho_0} r^{s-5/2},
\end{equation}
\begin{equation}
\dot{B_{\phi}}=\dot{B_0} r^{(s-5)/2},
\end{equation}
where $ \dot{\rho_0} $ and $ \dot{B_0} $ are constants.

Considering hydrostatic equation, we obtain the disc half-thickness $ H $ as
\begin{equation}
\frac{H}{r}=\frac{C_3 (1+\beta)}{2\frac{M_d}{M_*}}=\frac{C_3 (1+\beta)}{D},
\end{equation}
where $ M_d $ is the mass enclosed in the disc within a radius $ r $ and is given approximately by $ M_d=\pi r^2 \Sigma $. Also, we define parameter  $ D(=2\frac{M_d}{M_*} )$ as self-gravitating coefficient.

 By substituting the above similarity  solutions into the continuity and the motion equations and also the energy and the induction equations, we obtain the following system of dimensionless algebraic equations,
\begin{equation}
\dot{\rho}_0=-(s+\frac{1}{2})\frac{C_1 \alpha \Sigma_0 \sqrt{GM_*} D}{2(1+\beta) C_3},
\end{equation}
\begin{equation}
\frac{1}{2}\alpha^{2}C^{2}_{1}+C^{2}_{2}-1-\big[s-1+\beta(s+1)\big]C_3=0,
\end{equation}
\begin{equation}
C_1=\frac{3(s+1)(1+\beta)}{D}C^{3/2}_{3},
\end{equation}
\begin{equation}
\frac{H}{r}=\frac{(1+\beta)C_3}{D},
\end{equation}
\begin{equation}
C^{2}_{2} = \frac{2D}{9f(1+\beta)}(\frac{3-\gamma}{\gamma-1})C_1
C^{-1/2}_{3},
\end{equation}
\begin{equation}
\dot{B}_0= \frac{(3-s)}{2}\alpha C_1(GM_*)\sqrt{\frac{4 \pi D \Sigma_0
\beta}{(1+\beta)}},
\end{equation}
As is easily seen from the above equations, for $ s=-1/2 $ there is no mass loss, while there exist mass loss for $ s > -1/2 $. On the other hands, the escape and creation of magnetic fields are balanced each other for $ s = 3 $. Although outflow is one of the most important progresses in accretion theory,(see Narayan \& Yi 1995; Blandford \& Begelman 1999;   Stone \& Pringle \& Begelman 1999 and also some recent work like Xie \& Yuan 2008; Ohsuga \& Mineshige 2011), but in out model we chouse self-similar solution in a way that $\dot{\rho}=0$ and $ \dot{B}_{\varphi} \propto r^{-11/4}�$,$ (s = -1/2) $. So we ignored the effect of wind and outflow on the structure of the disks.

After some algebraic manipulations, we can reduce the above equations to a sixth order algebraic equation for $ C_1 $,
\begin{equation}
A^{3}C_1^{6}+3 A^{2}E C^{4}_{1}+( 3AE^{2}+B^{3}) C^{2}_{1}+E^{3}=0,
\end{equation}
where the coefficients depend on  the input parameter as
\begin{equation}
A= \frac{1}{2}\alpha^{2},
\end{equation}
\begin{equation}
B= \epsilon +\frac{1}{2}(3-\beta)(\frac{2D}{3(1+\beta)})^{2/3},
\end{equation}
\begin{equation}
E=-1,
\end{equation}
\begin{equation}
\epsilon = \frac{1}{3^{5/3}f}(\frac{2D}{1+\beta})^{2/3}(\frac{3-\gamma}{\gamma -1}).
\end{equation}
Having $C_1$ from this algebraic equation, the other variables (i.e. $C_2$ and $C_3$) can be determine easily.

\begin{figure}[t]
\input{epsf}
\epsfxsize=3.8in \epsfysize=3.2in
\centerline{\epsffile{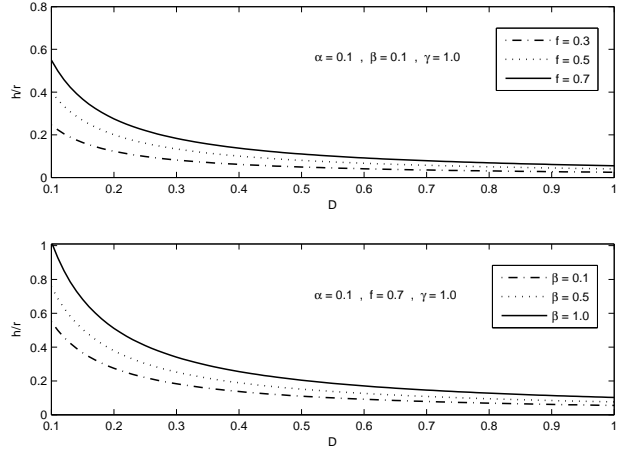}}
 \caption{These plots show behavior of the  $h/r$ versus  self-gravitating parameter $D$ for different values of the advection parameter $f$ upper panel and magnetic field parameter, $\beta$, lower panel. Here, the input parameters are  $ f=0.7 $, $ \alpha=0.1 $, $\beta=0.1$  and $ \gamma=1 $. }\label{c1-f_D}
\end{figure}

\begin{figure}[t]
\input{epsf}
\epsfxsize=3.8in \epsfysize=3.3in
\centerline{\epsffile{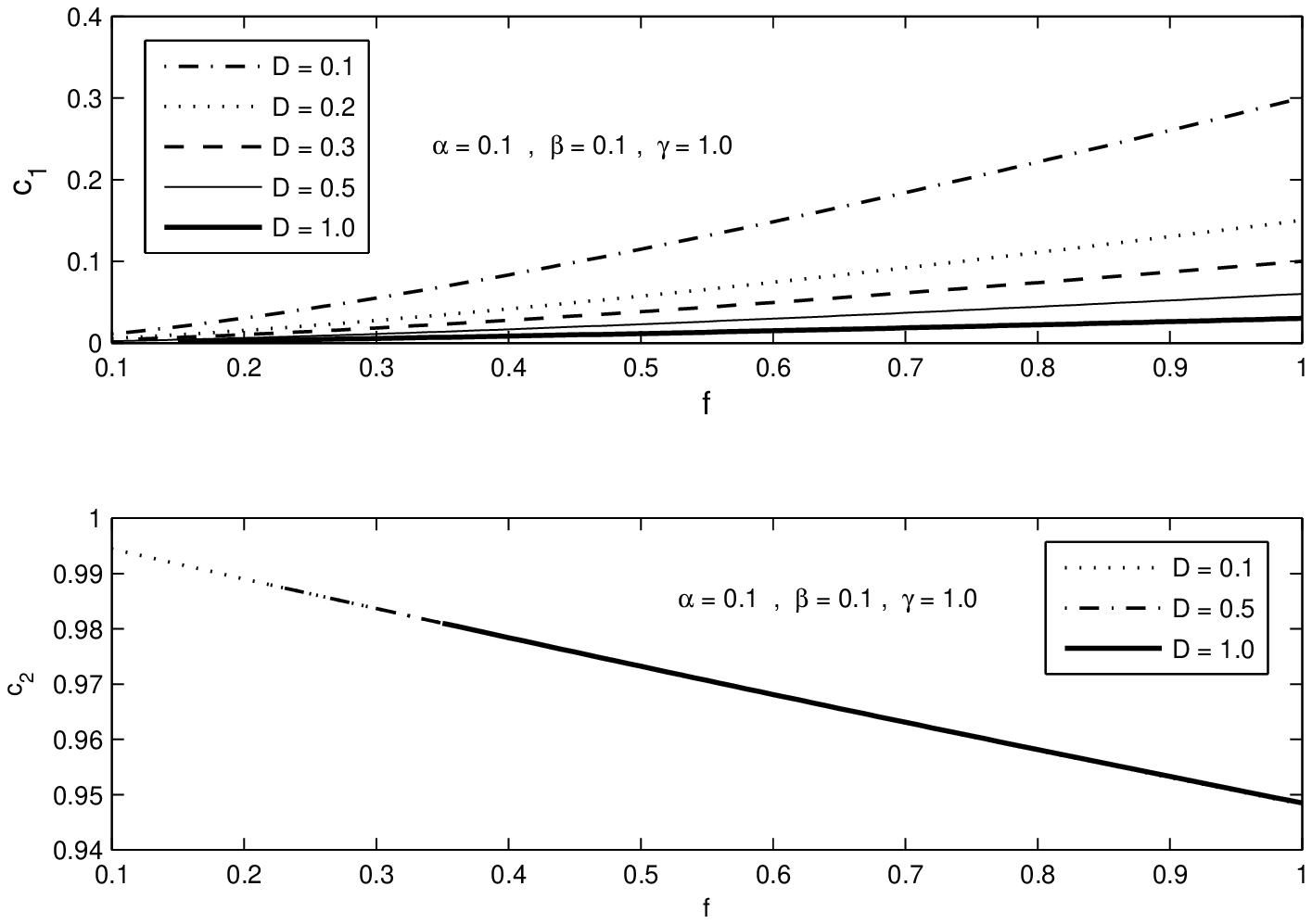}}
 \caption{These plots show behavior of the  coefficients $ C_1 (-\frac{v_r}{\alpha v_k}) $ and $C_2 (\frac{v_{\varphi}}{v_k})$ versus  advection parameter $f$ for different values of the self-gravitating parameter $D$. Here, the input parameters are  $ \beta=0.1 $, $ \alpha=0.1 $  and $ \gamma=1 $.  }\label{c2-f_alpha}
\end{figure}
\begin{figure}[t]
\input{epsf}
\epsfxsize=3.8in \epsfysize=3.2in
\centerline{\epsffile{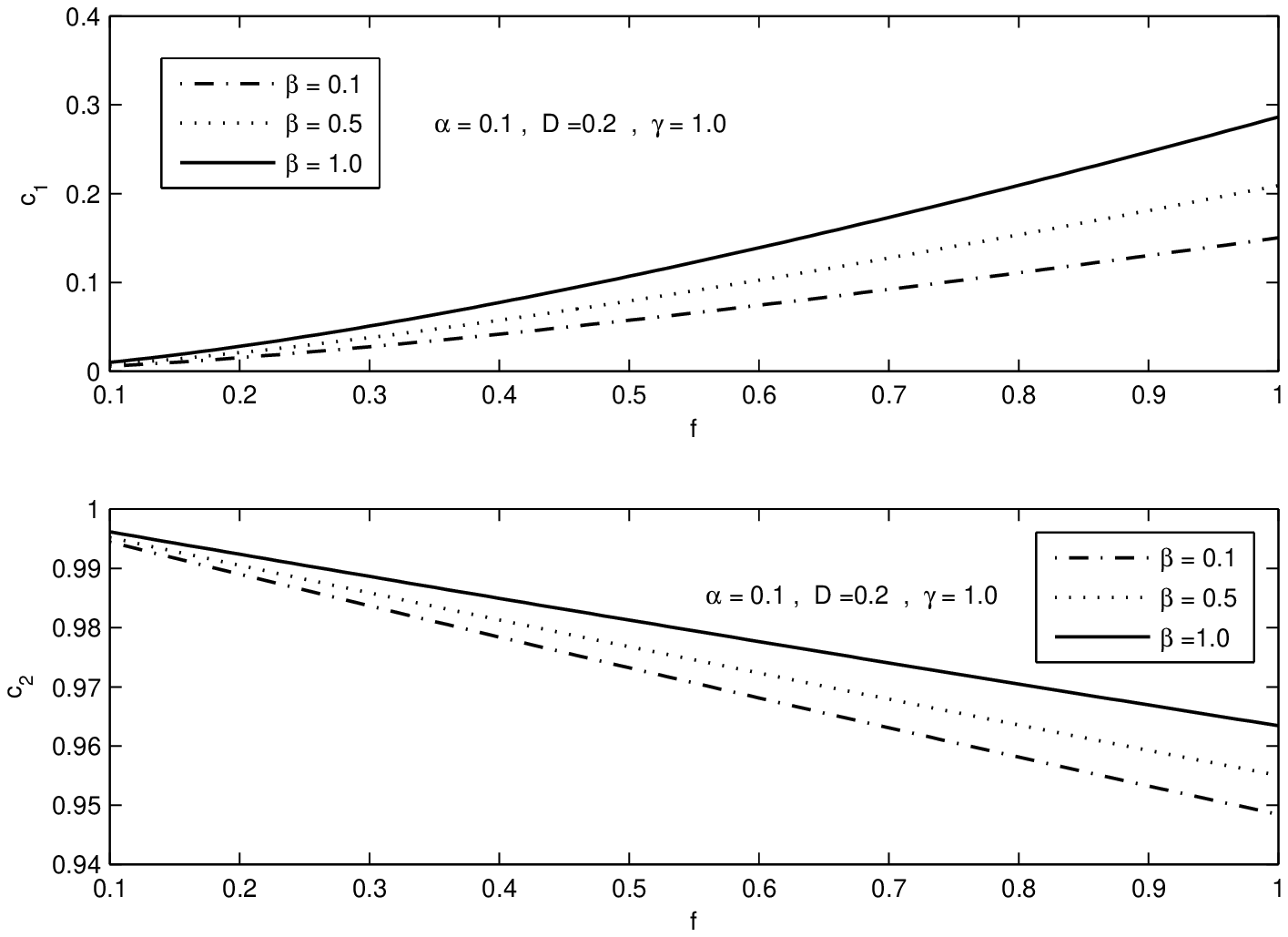}}
 \caption{These plots show behavior of the  coefficients $ C_1 (-\frac{v_r}{\alpha v_k }) $ and $C_2 (\frac{v_{\varphi}}{v_k})$ versus  advection parameter $f$ for different values of the magnetic field parameter $\beta$. Here, the input parameters are  $ D=0.2 $, $ \alpha=0.1 $  and $ \gamma=1 $.  }\label{c2-f1_alpha}
\end{figure}

\begin{figure}[t]
\input{epsf}
\epsfxsize=3.8in \epsfysize=3.4in
\centerline{\epsffile{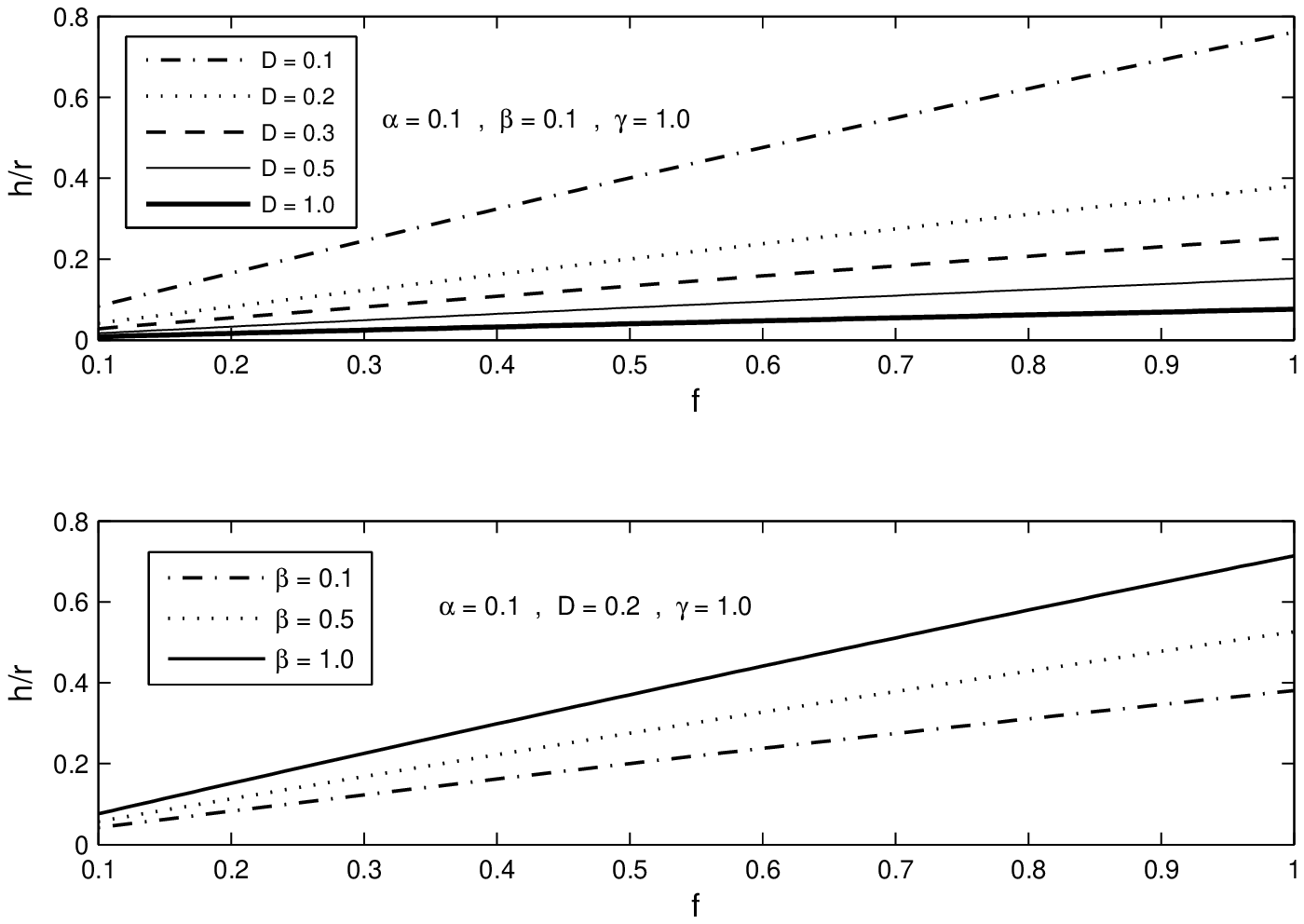}}
 \caption{These plots show behavior of the  $h/r$ versus  self-gravitating parameter $D$ for different values of the self-gravitating parameter $D$ upper panel and magnetic field parameter, $\beta$, lower panel. Here, the input parameters are  $ \beta=0.1 $, $ \alpha=0.1 $, $D=0.2$  and $ \gamma=1 $. }\label{c2-f_beta}
\end{figure}

 Now, we can do a parameter study considering our input parameters. Figure 1 shows how the coefficients $C_1$ and $C_2$ depend on the self-gravitation parameter $D$ for different values
of advection parameter $f$. Radial velocity is determined by $C_1$ which is shown in the upper panel. In ADAFs the radial velocity is generally less than free fall velocity on a point mass, but it becomes larger if
the advection parameter $f$ is increased and such a behavior   is consistent with the previous analytical solutions (e.g., Akizuki \& Fukue 2006;  Abbassi et al. 2008, 2010).
Obviously, larger parameter D implies a more self-gravitating disc and top plot of Figure 1 shows that the accretion velocity decreases as parameter $D$ increases. Compare to a non-self-gravitating disc, low $D$, the velocity of the flow will decrease some order of magnetude. Moreover, as more dissipated energy is advected within the flow, we see that the radial velocity is more sensitive to the variations of the self-gravitating parameter within the range  $D \preceq 5$. But when the disc becomes strongly self-gravitating (i.e., large $D$), the accretion velocity is more or less independent of the advected energy. Bottom plot of Figure 1 shows coefficient $C_2$ (i.e., rotational velocity) versus parameter $D$.  Here,  the rotational  velocity is not affected by the variation of the parameter $D$, but it  decreases when the  advection parameter $f$ increases.

 In Figure 2, we  assume the fraction of the advected energy is $f=0.7$ and then behavior of the  coefficients $C_1$ and $C_2$ versus parameter $D$ are shown for different values of magnetic field parameter $\beta$.
 For a given D, both the radial and rotational velocities increase as the disc becomes more magnetized (i.e., larger $\beta$) A magnetized disc must rotate faster than a case without
magnetic field because of the effect  of magnetic tension. The radial velocity is more sensitive to the variations of $D$ when the disc is more magnetized.  The rotational behavior is the same for different values of $D$ but it slightly shifts up when parameter $\beta$ increases.

In Figure 3, we show  vertical thickness of the disc in terms of the self-gravitating parameter $D$ for different values of $f$ and $\beta$.
The disc becomes thinner as the parameter $D$ increases. If we neglect force due to the self-gravity of the disc in the vertical direction, the disc becomes thicker as more energy advected with the flow. But our analysis shows for a strongly advected disc even not very large values of $D$ lead to a reduction of the disc thickness by a factor of two. For a given D, by increasing the parameters $f$ and $\beta$ the vertical thickness increases which means that advection and magnetic field will cause the disc becomes thicker.

Figure 4 shows how the coefficients $C_1$ and $C_2$ depend on the advection parameter $f$ for different values
of self-gravitating parameter $D$. Radial velocity is determined by $C_1$ which is shown in the upper panel. In the ADAFs solutions usually by adding advection parameter radial velocity will
increases while the rotational velocity decrease ( Akizuki \& Fukue 2006). Influence of the self-gravitation
of the disc will lead to decrease of radial velocity while the rotational velocity, $c_2$, is not affected. In Figures 5 and 6 the influence of
magnetic field strength, $\beta$, on the behavior of $C_is$ and vertical thickness of the flow were plotted respectively. By adding $\beta$, which indicates the role of magnetic field in the
dynamic of accretion discs, we will see that the radial flow increase as well as rotational velocities and vertical thickness increase. On the other hand radial and toroidal velocity increase when the
toroidal magnetic field becomes large. This is due to the magnetic tension term, which dominates
the magnetic pressure term in the radial momentum equation that assist the radial in-fall motion.

\section{Conclusions}

In this paper, we  studied an accretion disc in the advection
dominated regime considering purely  toroidal magnetic field and the
force due to the self-gravity of the disc in the vertical direction.
A set of similarity solutions was presented for such a
configuration. Our solutions reduce to the previous analytical
solutions (e.g., Akizuki \& Fukue 2006; Fukue 2004; and Abbassi et
al 2008) when self-gravity is neglected.  Some approximations were
made in order to simplify the main equations. We assume an axially
symmetric and static disc with $\alpha$-prescription for the
viscosity. We also ignored the relativistic effects.

Generally, it is believed that the ADAFs are hot and thick because of their inefficiency
in radiating out the dissipated energy. In particular, as more energy is advected,
the disc becomes thicker. But our analysis may slightly change this picture in particular
regarding to the thickness of the disc if the force due to the self-gravity of the disc
in the vertical direction is considered. Our disc still rotates with nearly Keplerian profile,
but its thickness is reduced significantly because of the self-gravity. In other words,
we may have a fully advected disc, but its thickness is reduced in comparison to a case without
self-gravity. Also, as the disc becomes more advective, the effect of self-gravity becomes more evident.
We think the present solutions are suitable for the outer parts of an ADAF not in inner parts.

It is difficult the evaluated the precise picture of Advection Dominated Accretion Flows (ADAFs) in the presence of B-field and self-gravity with self-similar method. However, this method can reproduce overall dynamical structure of the disc whit a set of given physical parameter. Although our preliminary self-similar solutions are too simplified, they clearly improve our understanding of physics of ADAFs around a black hole.

we believe the solutions are are presented, are applicable to the
discs with an inner hot part surrounded by an outer cold part where the self-gravity
plays an important role and and advection of the energy is not negligible. The discs of the
low-luminosity AGNs may have a similar situation and they do not show clear
transition from SSD to ADAF. The solutions are also applicable to
Slim discs, and thus have its potential applications in ULXs and
NLS1s.

It would be interesting if we add the effect of the wind and outflow in our future investigation to find out
how it would change solutions. The next step for future studies is to relax our simplifying assumptions, in
particular similarity method, and solve the relevant equations for
one of the mentioned systems and in doing so, our similarity
solutions will guide us. 

\acknowledgments
We are grateful to the referee for a very careful reading of the
manuscript and for his/her suggestions, which have helped us improve
the presentation of our results.

\end{document}